\documentclass[universe,article,submit,moreauthors,pdftex,10pt,a4paper]{mdpi} 
\usepackage{amssymb}
\firstpage{1} 

\def\tr{{\mathrm{tr}}}
\def\Hilbert{{\mathcal{H}}}
\def\H{{\mathrm{H}}}
\def\R{{\mathrm{R}}}
\def\E{{\mathrm{E}}}

\Title{Entropy budget for Hawking evaporation}

\Author{Ana Alonso-Serrano $^{1,2,\dagger,}$* and  Matt Visser $^{3,\dagger}$}
\AuthorNames{Ana Alonso-Serrano and Matt Visser}

\address{%
$^{1}$ \quad Institute of Theoretical Physics, Faculty of Mathematics and Physics, Charles University, 18000 Prague, Czech Republic\\
$^{2}$ \quad Instituto de F\'{i}sica Fundamental (IFF-CSIC), Serrano 121, 28006 Madrid, Spain; a.alonso.serrano@utf.mff.cuni.cz\\
$^{3}$ \quad School of Mathematics and Statistics, Victoria University of Wellington, PO Box 600, Wellington 6140, 
New~Zealand; matt.visser@sms.vuw.ac.nz}

\corres{Correspondence: a.alonso.serrano@utf.mff.cuni.cz}

\firstnote{These authors contributed equally to this work.}

\preto{\abstractkeywords}{\nolinenumbers}

\abstract{Blackbody radiation, emitted from a furnace and described by a Planck spectrum, contains (on average) an entropy of $3.9\pm2.5$ bits per photon. Since normal physical burning is a unitary process, this amount of entropy is compensated by the same amount of ``hidden information'' in correlations between the photons. The importance of this result lies in the posterior extension of this argument to the Hawking radiation from black holes, demonstrating that the assumption of unitarity leads to a perfectly reasonable entropy/information budget for the evaporation process. In order to carry out  this calculation we adopt a variant of the ``average subsystem'' approach, but consider a tripartite pure system that includes the influence of the rest of the universe, and which allows ``young'' black holes to still have a non-zero entropy; which we identify with the standard Bekenstein entropy.\vspace{2mm} \\ 
Based on a presentation at the conference: Varying Constants and Fundamental Cosmology 
(12-17 September 2016, Szczecin, Poland).\vspace{1mm}\\ Published as Universe 3 (2017) 58.}


\begin{document}


\section{Introduction}
Hawking's 1976 calculation \cite{Hawking}, of the thermal emission from a black hole, is often interpreted in terms of Clausius entropy, indicating that, starting from a star in some unknown pure state, after it collapses to a black hole and subsequently evaporates the system will (at the final stage)  be in a mixed state, with corresponding loss of information. This argument gave rise to the so-called black hole information paradox, and there exist very many very different proposals mooted to resolve it. (Such as: The information is irremediably lost, or it is stored in remnants or baby universes, or perhaps one has to appeal to the existence of new physical phenomena such as firewalls, fuzzballs, gravastars, etc.) Most of the current ideas are based on the maintenance of unitary, as in standard quantum mechanics, but in some situations this assumption gives rise to non-standard physical effects --- as in the case of the Page proposal \cite{Page-curve}, which motivated, in part, the idea of firewalls \cite{firewalls}.

\enlargethispage{20pt}
In order to obtain a better understanding of this problem, we first considered a standard unitary process,  standard thermodynamic burning, in order to show the exact quantity of entropy exchanged between the burning matter and the electromagnetic field, which (given unitarity) must be compensated for with information hidden in correlations between the photons involved in the process \cite{burning}. 
We have used this quite standard result as a starting point to understand what happens with the entropy/information budget in general relativistic black hole evaporation. In this context, we have constructed a specific model in which we have seen that there is  no paradoxical behaviour~\cite{Hawkingflux}. So, we claim that maybe the evaporation process is relatively benign.

\section{Entropy/information in blackbody radiation}
We begin by considering the standard thermodynamics process of burning matter, where it is well known that the underlying theory is unitary. (At least as long as one uses standard quantum mechanics.) Unitarity implies strict conservation of the von Neumann entropy. We will use this fact to correctly understand entropy budget when we calculate the entropy associated with the blackbody radiation.\footnote{There are alternative non-extensive generalizations of the classical Clausius entropy --- such as the Tsallis entropy. Classically the Tsallis entropy generalizes the Shannon entropy, and in a quantum situation it generalizes the von Neumann entropy. However considerations of the Tsallis entropy lie far beyond the scope of the present article --- for now, we are trying to understand the Hawking entropy budget using the standard tool of von Neumann entropy.}

The application of standard statistical mechanics to a furnace with a small hole leads to the notion of blackbody radiation. The reasoning that then gives rise to the Planck spectrum implies some coarse graining (that is, we choose to measure some aspects of the emitted photons and ignore others). In this process, every photon that escapes from the furnace transfers an amount entropy to the radiation field given by

\begin{equation} \label{E:entropy}
S=\frac{E}{T}=\frac{\hbar \omega}{T},
\end{equation}
where $E=\hbar \omega$ is the energy of the photon and $T$ is the temperature of the furnace. This is simply the Clausius definition of entropy. For convenience, from now on this single-photon definition of entropy will be measured in terms of bits, converted from ``physical'' entropy by means of the relation.

\begin{equation}
\hat{S}_2=\frac{S}{k_B \ln 2}.
\label{E:bits}
\end{equation}

Now take into account the effect of coarse graining the entropy, considering it in terms of the von~Neumann entropy, which is conserved under the evolution of the system.  The information hidden in the correlations hidden by the coarse graining process is simply

\begin{equation}
I_\text{correlations} = S_\text{coarse grained}-S_\text{before coarse graining}. 
\end{equation}

After these preliminary definitions, the next step is to calculate the average energy per photon in blackbody radiation (using the Planck distribution). We see

\begin{equation}
\langle E \rangle = \hbar \langle \omega \rangle= \frac{\int \omega f(\omega) d\omega}{f(\omega) d\omega}
=\frac{\pi^4}{30 \,\zeta (3)}\,k_BT,
\end{equation}
where $\zeta$ is the Riemann zeta function.
From this expression it is straightforward to calculate, using the definitions of equations (\ref{E:entropy}) and (\ref{E:bits}), the average entropy per blackbody photon. We find

\begin{equation}
\langle\hat S_2\rangle = \frac{{\langle E \rangle}}{k_B\,T\,\ln 2} =\frac{\pi^4}{30 \zeta (3) \ln 2}
\approx \; 3.897 \;  \hbox{ bits/photon}.
\end{equation}
The standard deviation, (simply coming from the fact that the Planck spectrum has a finite width), is

\begin{equation}
\sigma_{\hat S_2} 
 = \frac{\sqrt{\langle E^2 \rangle-\langle E \rangle^2}}{k_B\,T\,\ln 2}
 = \sqrt{\frac{10800\zeta(3)\zeta(5)}{\pi^8}-1 }  \;\; \frac{\pi^4}{30\, \zeta (3) \,\ln 2} 
\approx 2.522  \; \hbox{ bits/photon}.
\end{equation}
Overall, the average entropy per photon in blackbody radiation is~\cite{burning}:

\begin{equation} \label{Sbits}
\langle\hat S_2\rangle \approx 3.897 \pm 2.522 \;  \hbox{ bits/photon}.
\end{equation}
This expression is relevant when the only thing that we know about the photon is that it was emitted as part of some blackbody spectrum from a furnace at some (possibly unknown) temperature. The result depends only on the \emph{shape} of the Planck spectrum, the Clausius notion of entropy,  and quite ordinary thermodynamic reasoning.

\enlargethispage{40pt}
Since we know that the underlying physics is unitary, this entropy \emph{must} be compensated with an equal quantity of information. That information would be hidden in the photon-photon correlations that we did not take into account in our coarse graining procedure. The fact that even a standard unitary burning process has a precisely quantifiable entropy/information budget should not really come as a surprise, but it certainly does not seem to be a well-appreciated facet of quantum statistical mechanics.

\section{Hawking evaporation of black holes}
Taking into account the previous result, now we are interested in applying it to the study of the entropy budget associated with the evaporation process of a relativistic black hole. In order to develop our study we assume the existence of trapping or apparent horizons, which are enough for the Hawking radiation to exist \cite{apparent1,apparent2}, but which also allow the information to escape from the interior of the black hole. (Event horizons are not necessary for Hawking evaporation, and simply lead to unnecessary confusion.) Moreover we consider that the evaporation process is complete and the whole process unitary --- we shall then check if this assumption is consistent, and whether the information comes out throughout the Hawking radiation process. Specifically, the question is, how is entropy encoded in this process? Does the information emerge continuously or only at late stages of the evaporation process? (For instance, very late stages of the evaporation process, when the black hole mass, curvature, and Hawking temperature all become Planck scale, might best be viewed as a particle cascade, see for instance~\cite{Visser:ppp}.)
In order to understand the entropy/information budget we are going to calculate first the classical thermodynamic entropy and later the quantum (entanglement) entropy, in oder to compare and contrast them.

\subsection{Thermodynamic Clausius entropy in the Hawking flux}

\begin{figure}[!h]
	\begin{center}
		\includegraphics[scale=0.5]{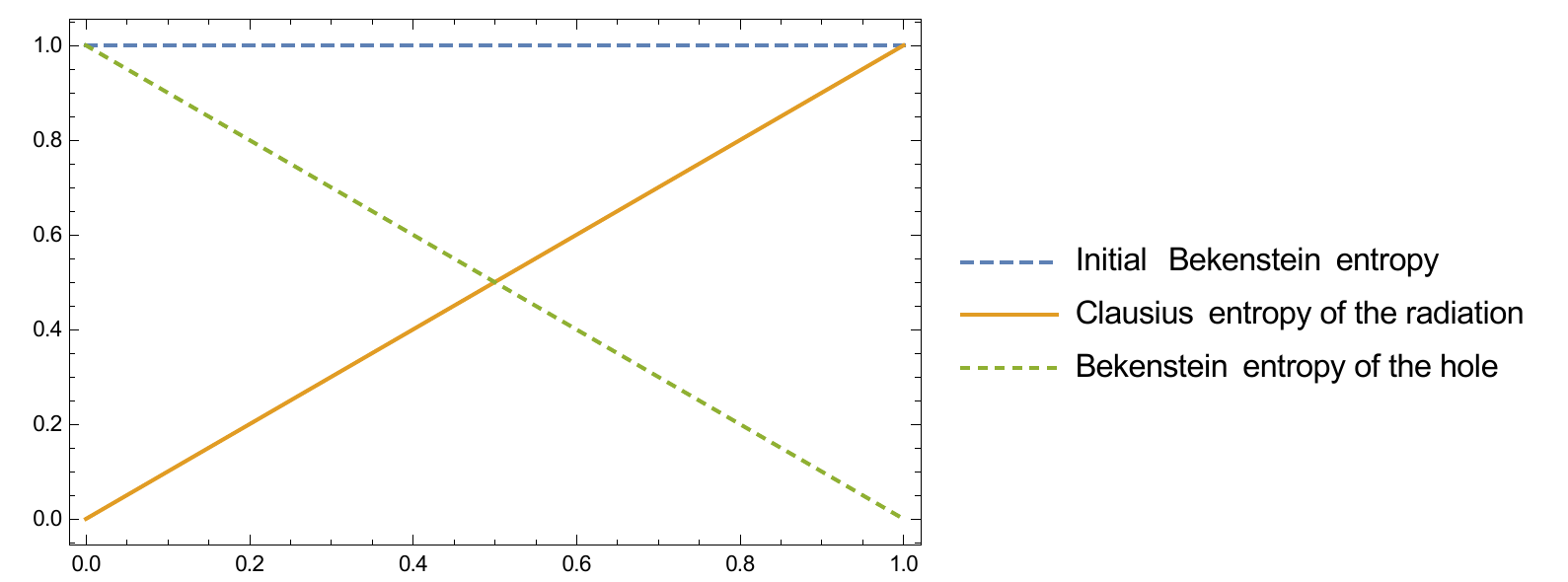}
		\caption{{\bf Clausius (thermodynamic) entropy balance:} \newline
			As the black hole Bekenstein entropy (defined in terms of the area of the horizon) decreases, the Clausius entropy of the radiation increases, to keep total entropy constant and equal to the initial Bekenstein entropy.}
		\label{F:entropy-balance}
	\end{center}
\end{figure}

The Bekenstein entropy \emph{lost} by a Schwarzschild black hole (per emitted quanta) is

\begin{equation}
{d S\over d N} ={d S/d t\over d N/d t} 
= {d(4\pi k_B G E^2/\hbar c)/d t\over  d N/d t }
= {8\pi k_B E} {dE\over dN} 
={1\over T_H} {dE\over dN} 
= {\hbar\langle\omega\rangle \over T_H} 
= {k_B  \pi^4\over30\;\zeta(3)}.
\end{equation}
In a similar way,  the Clausius entropy \emph{gain} of the external radiation field (per emitted quanta), that is, the entropy gain of the Hawking flux, is

\begin{equation}
{d S\over d N} = {d E/T_H\over d N} = {\hbar\langle\omega\rangle \,d N\over T_H \,d N} = {\hbar\langle\omega\rangle \over T_H} 
=  {k_B \pi^4\over30\;\zeta(3)}.
\end{equation}
Both entropies are the same (as they certainly should be) and their measure in bits will be exactly the same quantity that we calculated previously in the standard thermodynamic case [in equation~(\ref{Sbits})], ${d \hat S_2\over d N}=  {\pi^4\over30 \zeta(3)\ln 2 }\approx \;3.897 \;   \hbox{ bits/quanta}$.
So, it can be seen that throughout the whole evaporation process, semi-classically, we have:

\begin{equation}
\label{E:clausius}
S_\mathrm{Bekenstein}(t) + S_\mathrm{Clausius}(t) = \hbox{constant} = S_\mathrm{Bekenstein,0}.
\end{equation}
This is represented in figure (\ref{F:entropy-balance}), showing that semi-classically everything holds together very nicely.
(This should not be at all surprising since the normalization constant in the Bekenstein entropy was originally derived by integrating up the Clausius entropy in the Hawking flux.)

\subsection{Entanglement entropy in the Hawking flux}

In order to calculate the entanglement (von Neumann) entropy we have adopted a variant of a method developed by Page \cite{Page:subsystem} that allows us to calculate the average subsystem entropy of multipartite systems.
We take the Hilbert space of the total system to split as the tensor product of Hilbert subspaces corresponding to subsystems $A$ and $B$, (for now taking into account only a bipartite system), $\mathcal{H}_{AB} = \mathcal{H}_{A} \otimes \mathcal{H}_{B}$, and also take the total space to be in a pure state. So, we can define the subsystem density matrix of $A$ as $\rho_A = \tr_B( |\psi\rangle\langle\psi |)$, and $\rho_B = \tr_A( |\psi\rangle\langle\psi |)$ for subsystem $B$. Then, each subsystem has an associated subsystem von Neumann entropy, $\hat S_A= -\tr(\rho_A\ln\rho_A)$, and similarly $\hat S_B= -\tr(\rho_B\ln\rho_B)$. Indeed because the total system is in a pure state $\hat S_A = \hat S_B$. 

Taking a uniform average over all possible pure states of the total system,\footnote{This uniform averaging process always makes sense mathematically; when applied to black holes this mathematical process is a ``stand in'' for the only partially known physics that thermalizes both the Hawking radiation and the evolving black hole. Our initial input assumptions along these lines are certainly no stronger those used in deriving the Page curve.} the central result that Page obtained was that the subsystem entropy satisfies 

 \begin{equation}
 	\hat S_{n_1,n_2} = \langle \hat S_A \rangle = \langle \hat S_B \rangle  \lesssim \ln m,
 \end{equation}
 where $n_1=\dim(\Hilbert_A)$ and $n_2=\dim(\Hilbert_B)$ are the dimensions of the Hilbert spaces corresponding to each subsystem, and $m=\min\{n_1,n_2\}$ is the minimum of dimension of the two Hilbert spaces. Therefore, the subsystem entropy is very close to its maximum value, so that each subsystem is very close to being in a maximally mixed state. By combining the exact result derived by Sen \cite{Sen}, and the discussion carried out in reference \cite{Hawkingflux}, we can provide an strict limit to this entropy, given by 

 \begin{equation}
 	\hat S_{n_1,n_2} = \langle \hat S_A \rangle = \langle \hat S_B \rangle  
 	\in \left( \ln m - {\textstyle{1\over2}}, \ln m\right),
 \end{equation}
so the average subsystem is within $\frac{1}{2}$ nat of its maximum possible value.

\section{Bipartite entanglement}

The specific model considered by Page \cite{Page-curve} was a global system comprised of one subsystem that corresponds to the Hawking radiation and another subsystem that corresponds to the black hole, for which Hilbert spaces are given, respectively, by $\Hilbert_\mathrm{Hawking\,radiation} = \Hilbert_\R$ and $\Hilbert_\mathrm{black\,hole}= \Hilbert_\H$.
In this bipartite system, initially, before the evaporation of the black hole starts, there is not yet any Hawking radiation. Then, the Hilbert space $\Hilbert_\R$ is trivial, but the Hilbert space $\Hilbert_\H$ is enormous. 
(But note that one has to assume that the total system is in a pure state to apply Page's argument.) 
As the subsystem entropy is given by the minimum Hilbert space dimension, one has $(\hat S_{n_\H,n_\R})_0=0$, where the subscript indicates the initial state, $t=0$. 

In the opposite way, once the evaporation is completed, there will be no black hole, so, its Hilbert space dimension is trivial, and it is the Hawking radiation subspace which has an enormous Hilbert space dimension, giving $(\hat S_{n_\H,n_\R})_\infty=0$, where now the subscript indicates the final state, considered when $t=\infty$.

\begin{figure}[!h]
 	\begin{center}
 		\includegraphics[scale=0.5]{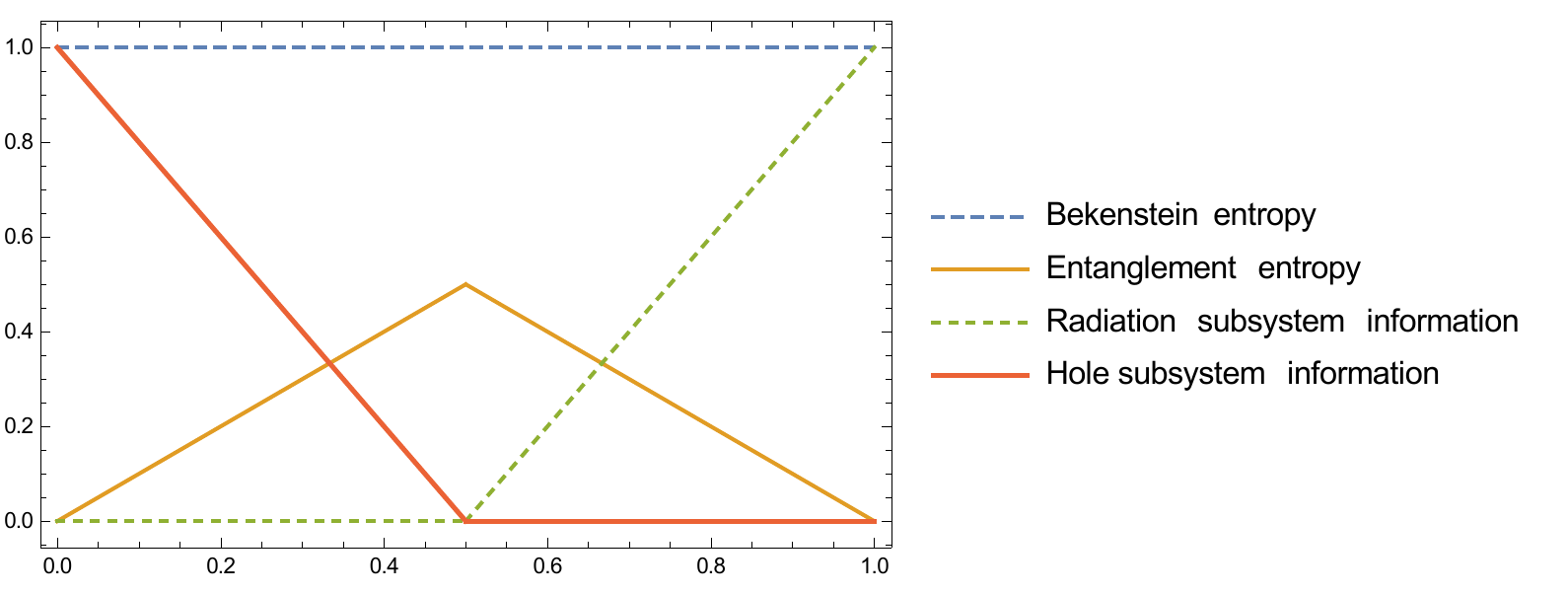}
 		\caption{{\bf Page curves for entanglement entropy and asymmetric subsystem information:} These are derived under the ``average subsystem'' assumption applied to a pure-state bipartite system consisting only of (black hole) plus the (Hawking radiation).}
 		\label{F:Page-curve}
 	\end{center}
 \end{figure}

In order to calculate the entropy at intermediate states, it is necessary to consider that the evolution is unitary, thus the total Hilbert space dimension is constant $	n_\H(t) \; n_\R(t) = n_\mathrm{H_0} = n_\mathrm{R_\infty}$. Under these conditions, the average subsystem entropy will be given in terms of

 \begin{equation}
 	\ln \min\left\{ n_\H(t),   {n_\mathrm{H_0}\over n_\H(t)} \right\}. 
 \end{equation}
 It is easy to find the maximum value of the average subsystem entropy, which is reached when $n_\H(t) \approx \sqrt{ n_\mathrm{H_0} }$, at which stage it takes the value
 
 \begin{equation}
 	\hat S_{n_\H,n_\R}(t=t_\mathrm{Page})\approx {1\over2} \; \ln  n_\mathrm{H_0}. 
 \end{equation}
 This time at which the black hole has lost half of its entropy is called the ``Page time''. It is possible to represent the shape of the evolution of this subsystem entropy, as it can be seen in figure (\ref{F:Page-curve}).  The so-called ``Page curve'' \cite{Page-curve} is the ``entanglement entropy'' curve.

\begin{figure}[!h]
 	\begin{center}
 		\includegraphics[scale=0.5]{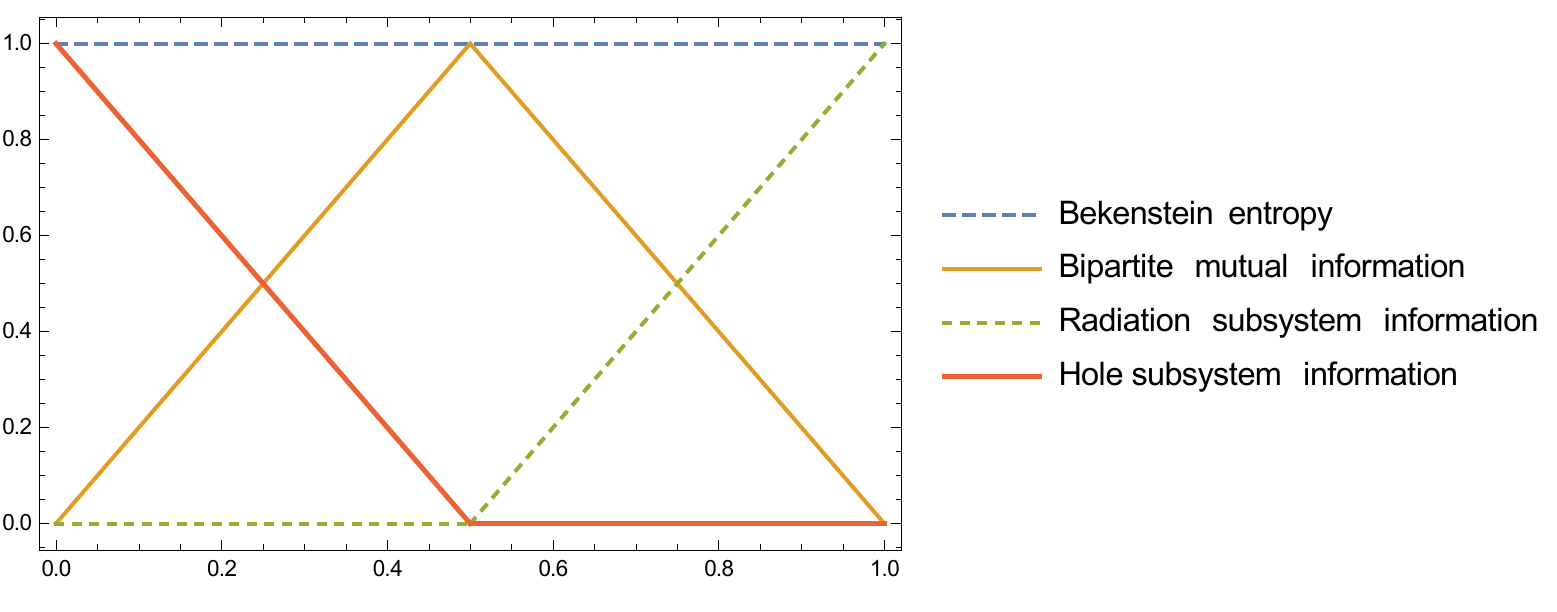}
 		\caption{{\bf Modified Page curves for the bipartite mutual information and the  asymmetric subsystem information:} These are derived under the ``average subsystem'' assumption applied to a pure-state bipartite system consisting only of (black hole) plus the (Hawking radiation).}
 		\label{F:Page-summary}
 	\end{center}
 \end{figure}

 \noindent
  Page also calculated the (averaged) asymmetric subsystem information given by the expressions~\cite{Page-curve} 
  
  \begin{equation}
  	\tilde I_{n_1,n_2}  = \ln n_1  - \hat S_{n_1,n_2},
  	\qquad
  	\tilde I_{n_2,n_1} = \ln n_2  - \hat S_{n_1,n_2};
  \end{equation}
 which are also represented in figure (\ref{F:Page-curve}). These are the curves labelled ``radiation subsystem information'' and ``hole subsystem information". 
In order to get a better understanding of the information budget, we have calculated the mutual entropy of the subsystems, given by

 \begin{equation}
 	I_{A:B} = S_A +S_B - S_{AB}.
 \end{equation}
 In this bipartite system it can be expressed as
 
 \begin{equation}
 	I_{\H:\R} = 2 S_\H = 2 S_\R.
 \end{equation}
  We have found that when we apply the ``average subsystem'' process to the mutual information, and combine it with the asymmetric subsystem information, we have that the ``sum rule'' is satisfied:
  
 \begin{equation}
 \langle \tilde I_{\H,\R} \rangle  + \langle \tilde I_{\R,\H} \rangle + \langle \hat I_{\H:\R}  \rangle \approx  
 	\ln\left(n_\H n_\R\right) \approx \hbox{constant} \approx \hat S_{\mathrm{Bekenstein,0}}.
 \end{equation}
 This sum rule is represented in figure (\ref{F:Page-summary}).

The Page curve underlies much of the present discussion about the ``information paradox''. The main result implies that the black hole subsystem is maximally entangled with the radiation subsystem. But, at the same time, if we sub-divide the Hawking radiation subsystem between early and late radiation (respectively, before and after Page time), these two subsystems would be also maximally entangled between them, and also with the black hole subsystem. The problem lies in the fact that, due to the monogamy of entanglement, this is not possible, and it was one of the motivations for the proposal of firewalls \cite{firewalls}. Nevertheless, we argue that this model misses much of the relevant physics.
 
\bigskip
There are some unacceptable  aspects of the standard argument. For instance, the standard argument assumes that the initial black hole (after formation but before any radiation is emitted) is in a pure state, so that the initial subsystem entropy vanishes. That assertion is in tension with the idea of relating the initial von Neumann entropy with the Bekenstein entropy of the black hole. That the Bekenstein entropy is a coarse-grained von Neumann entropy characterizing the number of ways in which the black hole could have formed is an old idea going back to the 1970s. Quantitatively, this idea was first formalized by Bombelli \emph{et al}.~\cite{Bombelli:1986rw}, and few years later was independently explored by Srednicki \cite{Srednicki:1993im}, both groups calculating the scaling of entanglement entropy with area (see, also, the  reviews~ \cite{Eisert:2008ur, Das:2008sy}, and the recent article on coarse-graining~\cite{Alonso-Serrano:coarse}). We propose that these problematic issues may be related to the consideration of an over-simplified (black hole)+(radiation) ``closed box'' system, ignoring the environment. That is, we argue that we should instead consider a tripartite system, in which we explicitly add the rest of the universe, (the environment), expecting a much better physically more reasonable behaviour for the entropy budget.

\section{Tripartite entanglement}
We now consider a tripartite system which consists of three subsystems,  associated to the black hole, the Hawking radiation, and rest of the universe (environment), respectively. So the Hilbert space now is split in the form: $\Hilbert_\mathrm{HRE}  = \Hilbert_\H \otimes \Hilbert_\R \otimes \Hilbert_\E$. Since we assume that the entire Universe is in a pure state ($S_{HRE}=0$), now the entropy of the subsystems is given by $S_\H(t) = S_\mathrm{RE}(t)$, $S_\R(t) = S_\mathrm{HE}(t)$, and $S_\E(t) = S_\mathrm{HR}(t)$.
In this case,\footnote{If the universe is not in a pure state the fix is simple: Divide the universe into observable and unobservable sectors. Take the entire universe to be a pure state and trace over the unobservable portion of the universe; the observable part of the universe is then described by a density matrix.  Discard the unobservable part of the universe (it is merely a spectator) and work with the black hole, the Hawking radiation, and rest of the observable universe (environment). The same argument goes through, and we see that for all practical purposes we might as well (without loss of generality) assume the universe is in a pure state.}  
the initial subsystem entropies, (before the evaporation starts, when $\dim(\Hilbert_\R)_0=1$), are $S_\mathrm{H_0} = S_\mathrm{E_0}$ and $S_\mathrm{R_0} = 0 = S_\mathrm{HE_0}$. It is important to realise that after black hole formation but before evaporation starts one has $S_\mathrm{H_0} = S_\mathrm{E_0} = S_{\mathrm{Bekenstein,0}}$.

Starting from any stellar object, collapse and horizon formation, (be it an apparent horizon, trapping horizon, event horizon, or some notion of approximate horizon), is an extremely dramatic coarse-graining processes.   We emphasize, (since we have seen this point cause some considerable confusion), that the end result of the collapse process is that the entropy of the newly formed black hole is the Bekenstein entropy associated with the horizon.  Indeed the Bekenstein entropy is the entropy associated with all possible ways the black hole could have been formed; not the entropy of the original stellar object that underwent collapse; that stellar entropy is unknown \emph{and unknowable} after black hole formation, this merely being one side effect of the ``no hair'' theorems.

(Indeed if one denies the applicability of Bekenstein entropy to the newly created black hole, then it is absolutely no surprise that one rapidly ties oneself up in logical knots when considering the Hawking emission process.)
 
More precisely: One can either appeal to Bekenstein's original papers to get (entropy) $\propto$ (area), and then fix the normalization constant using Hawking's original papers~\cite{Bekenstein0,Hawking0}. Alternatively, if you insist on working only with von~Neumann entropy, then one can use Srednicki's calculation showing that generically (entropy) $\propto$ (area) for any surface we cannot look behind~\cite{Srednicki:1993im}, and again fix the normalization constant using Hawking's original papers~\cite{Hawking0}. (See also Bombelli \emph{et al}'s calculations of the von Neumann entropy implied by the existence of a horizon~\cite{Bombelli:1986rw}.)

The final entropies, when the black hole is completely evaporated ($\dim(\Hilbert_\R)_\infty=1$) are \mbox{$S_\mathrm{H_\infty}=0= S_\mathrm{RE_\infty}$} and $S_\mathrm{R_\infty} = S_\mathrm{E_\infty}$.
We assume that the evolution is unitary, so the total Hilbert space is preserved. The environment does not participate directly in the evaporation process, since the role of the environment is merely to allow the initial $t=0$ black hole to have a nonzero entropy. After $t=0$ the environment evolves separately, that is, the unitary time evolution operator is the tensor product of a unitary operator corresponding to the environment and another unitary operator corresponding to the  black hole and Hawking radiation subsystems, $U_\mathrm{HRE}(t) = U_\mathrm{HR}(t) \otimes U_\E(t)$. Thus, the total Hilbert space of the environment and the total Hilbert space of the other two subsystems are independently preserved during the evaporation process, and we can express the conservation of the Hilbert space dimension as $n_\mathrm{E_0}=n_\E(t)=n_\mathrm{E_\infty} \equiv n_\E$ and $n_\H(t)\cdot n_\R(t) =   n_\mathrm{H_0}	= n_\mathrm{R_\infty}$.

\begin{figure}[!hb]
	\begin{center}
		\includegraphics[scale=0.5]{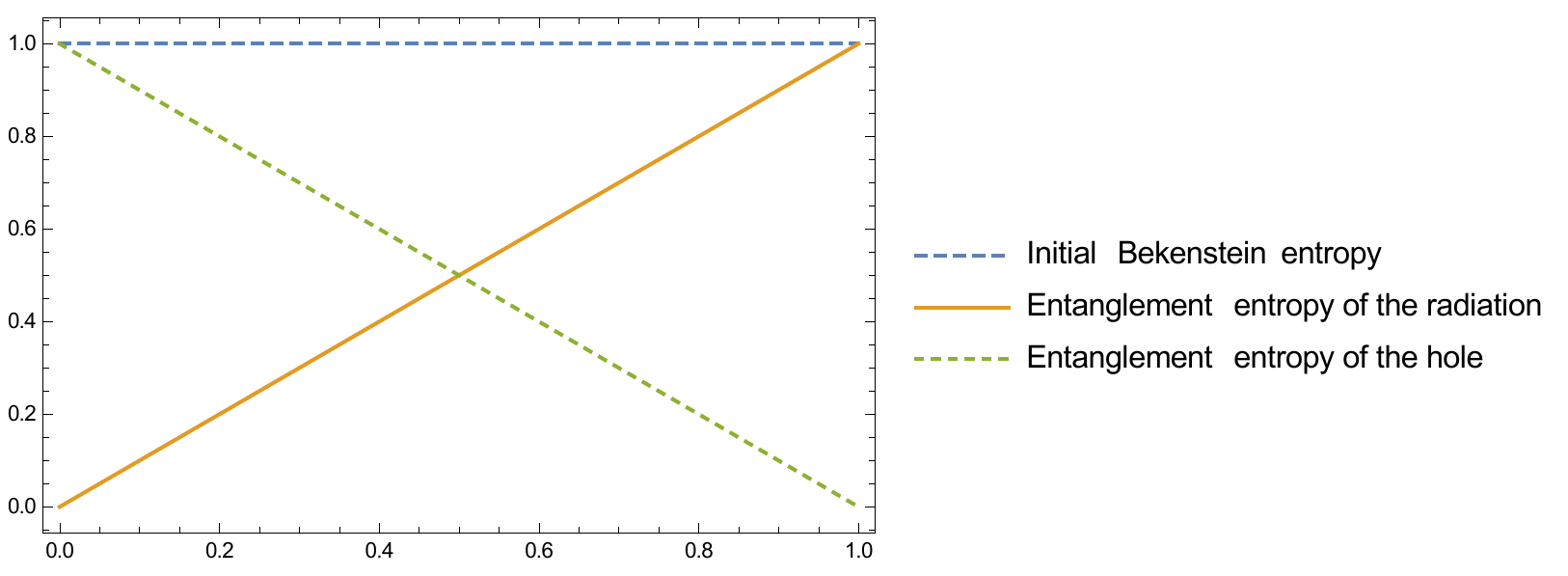}
		\caption{{\bf Tripartite quantum (von Neumann) entropy balance:} \newline Under the ``average subsystem'' assumption, now applied to a pure-state tripartite system consisting of (black hole) plus (Hawking radiation) plus (rest of universe), the quantum (von~Neumann) analysis reproduces the Clausius (thermodynamic) analysis. As the black hole Bekenstein entropy decreases, the entanglement entropy of the radiation increases, to keep total entropy approximately constant, at least to within 1 nat. In the limit where the environment (rest of universe) becomes arbitrarily large the correspondence is exact.}
		\label{F:entropy-balance2}
	\end{center}
\end{figure}

We have computed the average entropy of the black hole and Hawking radiation subsystems, taking as an additional assumption that (throughout the evolution) the Bekenstein entropy can be interpreted as the entanglement entropy of the black hole,

\begin{equation}
	\hat S_\mathrm{Bekenstein}(t) = \langle \hat S_\H(t)\rangle \approx 
	\ln\min\{n_\H(t), n_\R(t) \, n_\E \}\approx  \ln n_\H(t),
\end{equation}

\begin{equation}
	\langle \hat S_\R(t)\rangle \approx 
	\ln\min\{n_\R(t), n_\H(t) \, n_\E \}\approx 
	\ln n_\R(t).
\end{equation}
We have also obtained the sum of both averaged entropies. After some calculation

\begin{equation} \label{E:sum}
 \langle \hat S_\H(t)\rangle +\langle \hat S_\R(t)\rangle \approx  \ln [n_\H(t) \,n_\R(t)] = \hbox{constant} 
	= \ln n_\mathrm{H_0}.
\end{equation}
This sum rule is represented in figure (\ref{F:entropy-balance2}).

\noindent
In the same way we have obtained the average entropy of the environment subsystem. It is important to note that this entropy only corresponds to that part of the universe which is entangled with the other subsystems, it is not the entropy of the rest of the universe. Calculation  yields

\begin{equation}
	\langle \hat S_\E(t)\rangle=  \langle \hat S_\mathrm{HR}(t)\rangle\approx 
	\ln n_\mathrm{H_0} \approx \hbox{constant} \approx S_\mathrm{Bekenstein,0}.
\end{equation}
In this tripartite system, the mutual information between the Hawking radiation subsystem and the black hole subsystem is more interesting, and  is given by the expression

\begin{equation}
	I_{\H:\R} = S_\H +S_\R - S_{\H\R} = S_\H +S_\R - S_{\E}. 
\end{equation}
It is possible to calculate the average mutual information, and one finally finds \cite{Hawkingflux} that it is always less than $1/2$ nat during the whole  evaporation process,

\begin{equation}
	\langle \hat I_{\H:\R} \rangle \leq 
	{ n_\H n_\R \over 2 n_\E}  = { n_{\H_0}  \over 2 n_\E}  \leq {1\over2}.
\end{equation}
It is also interesting to note that if the environment becomes arbitrarily large, which is certainly possible in this tripartite system without any lost of generality, then it can be seen that the previous sum in equation (\ref{E:sum}) becomes exact

\enlargethispage{20pt}
\begin{equation}
	\lim_{n_\E\to\infty}  \left( \vphantom{\Big{|}} \langle S_\H \rangle +  \langle S_\R \rangle \right)
	= \lim_{n_\E\to\infty}  \langle S_\E \rangle,
\end{equation}
and the mutual entropy is also exactly zero \cite{Hawkingflux}

\begin{equation}
	\lim_{n_\E\to\infty} \langle I_{\H:\R} \rangle = 0.
\end{equation}

\section{Discussion}
First of all, we have obtained the numerical value of the entropy per photon emitted in black body radiation, which, because of the process is unitary, must be compensated by an equal “hidden information” in the correlations. As is well known there is no ``information puzzle'' in a standard thermodynamic process, but we note that due to the coarse-graining~\cite{Alonso-Serrano:coarse},  a specifically quantifiable amount of entropy/information is nevertheless exchanged in the process~\cite{burning}.

From this starting point, we have applied these ideas to the consideration of general relativistic black holes, calculating both the classical thermodynamic entropy and the Bekenstein entropy, and seeing that they compensate perfectly. Once we have calculated the classical entropy, we then calculate the quantum (entanglement) entropy, considering a model based on a tripartite system. The result obtained  is completely in agreement with the classical expected results, at least to within 1 nat. In contrast, the result previously obtained by Page, by considering a bipartite model that does not interact with the environment, gives rise to not well-understood physics.

From our analysis, it can be seen that although when we restrict attention to any particular subsystem we perceive an amount of entanglement entropy, (a loss of information), there exists a complementary amount of entropy/information that is codified in the correlations between the subsystems. Then, assuming the unitarity of the evolution of the (black hole) + (Hawking radiation) subsystem, and working within the standard Page-like average-subsystem framework, we showed that it seems that there is no pressing need for any unusual physical effect to enter into the process. This implies a continuous purification of the Hawking radiation, and could lead to a completely non-controversial and quite standard physical picture for the evaporation of a black hole. (Here we are taking into account only the semiclassical process of Hawking radiation,  until a deeper understanding of the underlying micro-physics of quantum gravity phenomena might be found).

\vspace{6pt} 


\acknowledgments{AA-S is supported by the grant GACR-14-37086G of the Czech Science Foundation. \\
MV is supported by the Marsden fund, administered by the Royal Society of New Zealand.}

\authorcontributions{Both authors contributed equally to this article.}

\conflictofinterests{The authors declare no conflict of interest.} 
\bibliographystyle{mdpi}

\renewcommand\bibname{References}



\end{document}